\documentclass{Interspeech2024}
\newcommand\dv{\text{d}}
\usepackage{xcolor}
\usepackage{soul}
\usepackage{textcomp}  
\usepackage{scalerel}  

\def\�{\scalerel*{\includegraphics{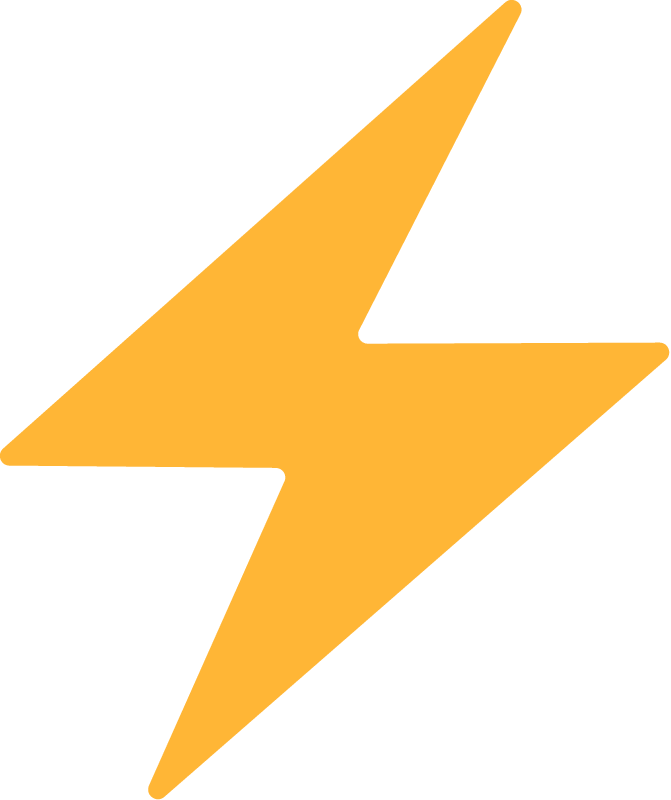}}{\textrm{\textbigcircle}}}

\newcommand{\cmark}{\ding{51}}%
\newcommand{\xmark}{\ding{55}}%
\interspeechcameraready 

\title{Thunder\� : Unified Regression-Diffusion Speech Enhancement with a Single Reverse Step using Brownian Bridge}

\name[affiliation={1}]{Thanapat}{Trachu}
\name[affiliation={2}]{Chawan}{Piansaddhayanon}
\name[affiliation={1, 2}]{Ekapol}{Chuangsuwanich}

\address{
  $^1$Department of Computer Engineering, Faculty of Engineering, Chulalongkorn University \\ $^2$Center of Excellence in Computational Molecular Biology, Chulalongkorn University} 
\email{thanapat.trachu@gmail.com, schwanph@gmail.com, ekapolc@cp.eng.chula.ac.th}

\begin{document}
\maketitle
\ninept
 
\begin{abstract}

Diffusion-based speech enhancement has shown promising results, but can suffer from a slower inference time. Initializing the diffusion process with the enhanced audio generated by a regression-based model can be used to reduce the computational steps required. However, these approaches often necessitate a regression model, further increasing the system's complexity. We propose \textit{Thunder}, a unified regression-diffusion model that utilizes the Brownian bridge process which can allow the model to act in both modes. The regression mode can be accessed by setting the diffusion time step closed to 1. However, the standard score-based diffusion modeling does not perform well in this setup due to gradient instability.  To mitigate this problem, we modify the diffusion model to predict the clean speech instead of the score function, achieving competitive performance with a more compact model size and fewer reverse steps.

\end{abstract}

\section{Introduction}

Speech enhancement (SE) focuses on removing noisy signals from the input speech to improve its comprehensibility and has been deployed in several real-life systems \cite{app10176077, 9414346, shon2019voiceid}. It could also be integrated with existing downstream tasks, such as speech recognition (ASR) \cite{zhu2022joint, maas2012recurrent, koizumi2021snri} or speech verification (SV) \cite{shon2019voiceid, eskimez2018front}, to improve speech quality under adverse environments.

Speech enhancement systems, as categorized in \cite{StoRM}, can be classified into two approaches: regressive \cite{conv_tasnet, metricgan+, gagnet} and generative \cite{segan, CDiffSE, SGMSE+}\footnote{In this paper, a regressive model is a deterministic mapping between noisy and clean speech while the generative model is not.}. The objective of the regression model is to learn a deterministic mapping between noisy and clean speech, whereas the generative model aims to capture the target distribution, allowing the generation of multiple valid possibilities instead of a single one. 
Recently, there has been a surge of interest in the diffusion model for speech enhancement \cite{CDiffSE, SGMSE+} due to its promising outcomes across various domains \cite{ScoreBasedDiffusion}.




Despite the promising outcome, one major obstacle to the practical application of diffusion for SE is its slow inference time caused by multiple reverse diffusion steps. Thus, numerous studies have been proposed to address this issue. StoRM \cite{StoRM} utilized a two-stage regression-diffusion pipeline where the first model is responsible for enhancing the noisy speech in a regressive manner while the second stage is used for refining the output from the former stage using a reverse diffusion process. Since the input to the latter model is pre-cleaned, the difficulty of the reverse process decreases, requiring fewer diffusion steps. Nevertheless, this approach requires two independent models—regression and diffusion—leading to a substantial increase in the number of parameters. To address this issue, an additional head to predict both the score function and the noiseless signal was introduced in the Diffusion-based Joint Predictive and Diffusion model \cite{DiffusionJointPredictive}, achieving competitive outcomes while incurring fewer parameters. Nevertheless, it still requires an additional prediction head for regressive prediction.

 We introduce Thunder, a unified regression-diffusion model capable of performing both regression and diffusion while not incurring additional parameters. We propose the use of the Brownian bridge process for diffusion-based speech enhancement which allows the model to act as both a regression and a diffusion model at the same time. Instead of modeling the score function like in typical diffusion modeling, we reparameterize the model to predict the noiseless speech to avoid the gradient instability issue and allow a single step prediction if desired (regression mode).  Our method achieves competitive results on the VoiceBank + DEMAND dataset using fewer parameters and shorter inference time. Remarkably, our approach outperforms the diffusion baselines on even just one reverse diffusion step highlighting the effectiveness of the Brownian bridge process. 




\section{Score-based diffusion model}
\subsection{Forward and reverse process}
Diffusion modeling comprises two essential processes: the forward process and the reverse process. In the forward process, noise is incrementally introduced into a clean speech until it becomes pure noise. Conversely, the reverse process gradually eliminates noise from noisy speech, ultimately yielding clean speech. Within the framework of score-based diffusion \cite{ScoreBasedDiffusion}, a stochastic differential equation (SDE) is employed to represent these processes. Specifically, the forward process is represented by the following SDE:
\begin{align} \label{eq:forward_process}
    \dv x_t = f(x_t, y) \dv t  + g(t) \dv w
\end{align}
where $x_t, y, w$ denotes the current state of the process at time step $t$, noisy speech, and a standard Wiener process, respectively. The state $x_t$ is indexed by a continuous time variable $t$ within the interval $[0, 1]$, in which $x_t$ is a clean speech when $t=0$ and a pure noise when $t=1$.
The functions $f(x_t, y)$ and $g(t)$ signify the drift coefficient and diffusion coefficient, respectively.
Following \cite{ScoreBasedDiffusion}, the reverse SDE of the Eq. \ref{eq:forward_process} is:
\begin{align} \label{eq:reverse_process}
    \dv x_t = [f(x_t, y) - g(t)^2 \nabla_{x_t} \log p_t(x_t)] \dv t + g(t) \dv w
\end{align}

There have been works \cite{CDiffSE, SGMSE+} proposed to design the SDE process for speech enhancement tasks by designing $f(x_t, y)$ and $g(t)$ that could directly transform the noisy speech into clean speech instead of Gaussian noise. For example, SGMSE+ \cite{SGMSE+} proposed the following drift and diffusion coefficient:
\begin{align}
    f(x_t, y) &= \gamma(y-x_t)\\
    g(t) &= \sigma_{\text{min}} (\frac{\sigma_{\text{max}}}{\sigma_{\text{min}}})^t \sqrt{2 \log (\frac{\sigma_{\text{max}}}{\sigma_{\text{min}}})}
\end{align}
where $\gamma$ denotes the transformation speed between the clean speech and the noisy speech, and $\sigma_\text{min}, \sigma_\text{max}$ are the parameters controlling the variance in $x_t$. However, the presences of Gaussian noise still exist at $t=1$ due to a non-zero variance. Therefore, in this paper, we have selected the subsequent drift and diffusion coefficients:
\label{eq:3}
\begin{equation}
f(x_t, y) = \frac{y - x_t}{1 - t}; \quad g(t) = 1
\end{equation}

This particular SDE is referred to as the Brownian bridge process \cite{BrownianBridgeProcess}. Its distinguishing feature is that it can linearly transform between the initial state $(x_0)$ with zero variance to the noisy speech $y$ with zero variance, offering a capability to perform as a regression model at $t=1$ (deterministic mode). 

\subsection{Score-function model}

As calculating $\nabla_{x_t} \log p_t(x_t)$ is intractable, following \cite{ScoreBasedDiffusion, ScoreMatching}, \textit{denoising-score-matching} is instead performed by having the score-based model $s_\theta(x_t, y, t)$, typically a neural network, approximates $\nabla_{x_t} \log p_t(x_t|x_0)$, the value of which can be determined using the given initial state $x_0$ \cite{StochasticDifferentialBook}:
\begin{align}
p_t(x_t|x_0, y) &= \mathcal{N}_\mathbb{C} (x_t; \mu(x_0, y, t), \sigma(t)^2 \mathbf{I}) \label{eq:prob_x_t}\\
\mu(x_0, y, t) &= x_0 (1-t) + yt \\
\sigma(t)^2 &= t(1-t) \label{eq:sigma}
\end{align}
where $\mathcal{N}_\mathbb{C}$ represents the circularly symmetric complex normal distribution, $\mu(x_0, y, t)$ denotes a mean, and $\sigma(t)$ is a standard deviation. 
Consequently, the training loss is defined as:
\begin{align} \label{eq:objective}
\mathcal{J}(\theta) = \mathbb{E}_{t, x_t, (x_0, y) \sim p_\text{data}} [\lambda(t)||s_\theta(x_t, y, t) + \frac{z}{\sigma(t)}||_2^2]
\end{align}
where $\mathcal{J}(\theta)$ is an objective function, and $t$, $x_t$ are randomly sampled from $\mathcal{U}[0, 1]$ and $p_t(x_t|x_0)$, respectively. $z$ is drawn from $\mathcal{N} (0, \mathbf{I})$, and $\lambda(t)$ serves as a weight function that is set to $\sigma(t)^2$ in \cite{StoRM, SGMSE+, ReducePrior}. 


\subsection{Inference} \label{section:inference}
To generate the predictions, the reverse SDE has to be estimated through a numerical SDE solver using the PC sampler \cite{ScoreBasedDiffusion} consisting of a predictor and corrector.
  Initially, $x_1$ is set to $y$. Then, the predictor updates the current state $x_t$ into the next state $x_{t-\Delta t}$ by discretizing the reverse SDE using finite time steps that is subsequently fed to the corrector to refine the prediction by using only the score function. The process was iteratively repeated until $t=0$. In this paper, we follow \cite{ScoreBasedDiffusion} and use the Euler-Maruyama and Langevin dynamics as a predictor and corrector, respectively.


\section{Methodology}

Drawing inspiration from StoRM \cite{StoRM} and the Joint Generative and Predictor method \cite{DiffusionJointPredictive}, we propose to further condense StoRM into a single model that can switch between two modes: diffusion and regression. Specifically, we train the model to predict $x_0$ instead of the score function and leverage the property of the Brownian bridge process to enable regressive capability.

\subsection{Model parameterization} 
\label{model_parameterization}

To allow the model to possess a regressive capability, the Brownian bridge process is employed. However, directly applying this process to the SDE is inappropriate since $\sigma(t)$ becomes very close to 0 when $t\to1$ (Eq. \ref{eq:sigma}), making minimizing the Eq. \ref{eq:objective} impractical as the gradient is directly proportional to $\sigma(t)$, as shown below.
\begin{align} 
\nabla_\theta \mathcal{J}(\theta) &= \nabla_\theta[||\sigma(t) s_\theta(x_t, y, t) + z||_2^2] \\
&= 2\sigma(t) \nabla_\theta s_\theta ||\sigma(t)s_\theta(x_t, y, t) + z||_2 \label{eq:gradient}
\end{align}
This hampers the model's ability to efficiently estimate the score function at $t=1$ under one reverse step (Eq. \ref{eq:reverse_process}). Even if the accurate score function $s_\theta$ is to be obtained, it is still infeasible to employ the regression mode at $t=1$ due to the inability to estimate the clean speech ($x_0$) from the score function as shown in the following equations, derived from Eq. \ref{eq:prob_x_t}:
\begin{align} \label{eq:9}
    x_t &\sim \mathcal{N}_\mathbb{C} (\mu(x_0, y, t), \sigma(t)^2 \mathbf{I}) \\
    x_t &= x_0 (1-t) + yt + \sqrt{t(1-t)} z \label{eq:reparameterization} \\ 
    x_t &= x_0 (1-t) + yt - t(1-t) s_\theta(x_t, y, t) \label{eq:sampling_x_t_from_score_function}\\
    x_0 &= \frac{x_t - yt + t(1-t)s_\theta(x_t, y, t)}{1-t}
\end{align}
where Eq. \ref{eq:reparameterization} follows the reparameterization trick from \cite{vae}, and $s_\theta (x_t, y, t) \approx -z / \sigma(t)$ when optimal.


To overcome this problem, we modify the model to predict $\tilde{x}_\theta(x_t, y, t)$, an estimation of clean speech $x_0$, instead of the score function, allowing our model to be used as a regression model at any $t$. 
In diffusion mode, we can perform the reverse process by first computing the score function via: 
\begin{align} \label{eq:clean_speech_to_score}
    s_\theta(x_t, y, t) &= -\frac{x_t - (\tilde{x}_\theta(x_t, y, t)(1-t) + yt)}{t(1-t)}
\end{align}
Then, the obtained score function can be used to solve the reverse SDE as described in 2.1.3.
During inference, at the initial stage of the reverse process, $t$ is set close to 1 to circumvent numerical instability. The training approach remains the same, with the training objective adjusted to: 
\begin{align}
\mathcal{J}(\theta) = \mathbb{E}_{t, x_t,  x_0, y} [||\tilde{x}_\theta(x_t, y, t) - x_0||_2^2]
\end{align}

\subsection{Justification for the Brownian bridge process} \label{bb_property}
This subsection provides some analysis to justify our choice of the Brownian bridge process, the drift and diffusion coefficient of the SDE, for the speech enhancement task. The \textit{drift coefficient} of the reverse Brownian Bridge as $t\to1$ converges to the noise in the speech as shown in the equations below:
\begin{align} \label{eq:score_function_in_form_of_x_0}
&  \lim_{t \to 1} f(x_t, y) - g(t)^2s_\theta(x_t, y, t) \nonumber \qquad (\text{From } (2))\\
& = \lim_{t \to 1} \frac{t(y-x_t)+ x_t - (\tilde{x}_\theta(x_t, y, t)(1-t) + yt)}{t(1-t)} \nonumber \\
& = \lim_{t \to 1} \frac{(1-t)x_t - \tilde{x}_\theta(x_t, y, t)(1-t)}{t(1-t)} \nonumber \\
& = \lim_{t \to 1} \frac{x_t - \tilde{x}_\theta(x_t, y, t)}{t} \nonumber \\
& = y - \tilde{x}_\theta(x_1, y, 1) = \tilde{n}
\end{align}
The equation above suggests that the reverse process would update in the direction of noise in the input speech $\tilde{n}$, matching the modeling assumption of speech enhancement where the noisy speech is the clean speech corrupted by the noise through an additive operation ($y=x_0+n$).
In an extreme scenario where the number of steps in the reverse process is set to 1, the model transforms into the regression mode, deterministically updating $x_1$ by subtracting the predicted noise. By enabling the regressive ability, the model could now enhance the speech with fewer reverse diffusion steps when initialized with the enhanced audio from the regression mode \cite{StoRM}.

\subsection{Utilizing the regression potential of the score-based model}

\begin{figure}[t!]
  \centering
  \includegraphics[width=.9\linewidth]{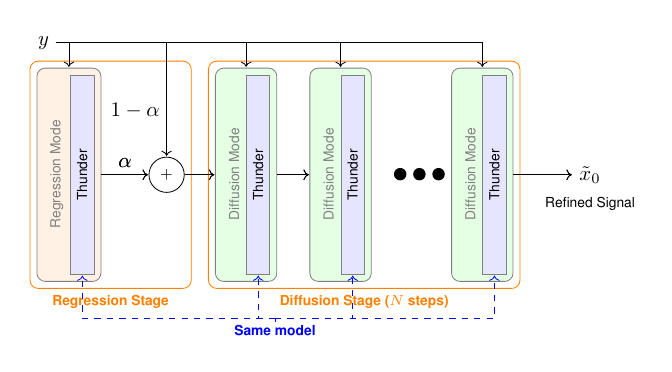}
  \caption{A summarization of Thunder during inference. The regression mode is first applied to the noisy input speech to improve the signal quality before being further refined through the diffusion mode. To reduce over-denoising artifacts caused by the regression part, the processed signal is fused with the original input to preserve its characteristic. The weights are shared across the two modes.
  }
  \label{fig:flowchart}
\end{figure}

 Figure \ref{fig:flowchart} summarizes the inference process. Following StoRM \cite{StoRM}, our pipeline is a two-stage process which is a regression model followed by a generative model. The first improves the signal quality, while the latter aims to reduce the artifacts generated by the regression model. However, unlike StoRM, the two models are shared but used slightly differently in different modes. The regression mode is done by predicting $x_0$ directly from our model $\tilde{x}_\theta(x_1 = y, y, t=1)$. The diffusion mode is performed by acquiring the score function according to Eq. \ref{eq:clean_speech_to_score}, and the reverse diffusion process can be performed for $N$ steps as outlined in Section \ref{section:inference}.
 
 As suggested in \cite{ArtifactsASR}, the input to the diffusion stage is a linear interpolation between the output from the regression mode and the noisy input speech which can help minimize the occurrence of the \textit{over-denoising artifact}, a concept discussed and employed in \cite{StoRM, CDiffSE, SGMSE}. The interpolation weight $\alpha$ can be chosen via grid search on a validation set.

\setlength\tabcolsep{3pt}
\begin{table}[t]
  \caption{Performance of different speech enhancement methods on the VoiceBank + DEMAND dataset. ``Type'' refers to the model type (``R'' for regression and ``G'' for generative model). Numbers before and after slash refer to the performance of small (S) and large (L) NCSN++ variants, respectively.
  The model with the best performance in each section is underlined, and the best score in the table is bolded.}
  \label{tab:performance}
  \centering
  \begin{tabular}{l c c c c}
    \toprule
    \textbf{System} & \textbf{Type} & \textbf{PESQ} $\uparrow$ & \textbf{ESTOI} $\uparrow$ & \textbf{SI-SDR} $\uparrow$\\
    \midrule
    Noisy & - & 1.97 & 0.79 & 8.4 \\
    \midrule
    Conv-Tasnet \cite{conv_tasnet} & R & 2.84 & 0.85 & 19.1 \\
    MetricGAN+ \cite{metricgan+} & R & \underline{\textbf{3.13}} & 0.83 & 8.5 \\
    NCSN++M (L) \cite{analysing_discriminative_diffusion} & R & 2.82 & \underline{0.87} & \underline{\textbf{19.9}} \\
    \midrule
    SEGAN \cite{segan} & G & 2.16 & - & - \\
    CDiffuSE \cite{CDiffSE} & G & 2.46 & 0.79 & 12.6 \\
    SGMSE+ (L) \cite{SGMSE+} & G & 2.93 & 0.87 & 17.3 \\
    BBED (L) \cite{ReducePrior} & G & 2.95 & 0.87 & 18.7 \\
    StoRM (S) \cite{StoRM} & R+G & 2.93 & \underline{\textbf{0.88}} & \underline{18.8} \\
    GP-Unified (L) \cite{DiffusionJointPredictive} & R+G & \underline{2.97} & 0.87 & 18.3 \\
    \midrule
        \textbf{Thunder (S/L)}& &  & &  \\
    Regression mode & R & 2.78/2.85 & 0.87/0.87 & \underline{19.6}/\underline{19.7} \\
    Diffusion mode & G & 2.87/2.95 & 0.87/0.87 & 18.8/18.6 \\
    Mixture ($\alpha=0.8$) & R+G & \underline{2.97}/\underline{3.02} & \underline{0.87}/\underline{0.87} & 19.3/19.4 \\
    \bottomrule
  \end{tabular}
\end{table}

\setlength\tabcolsep{3pt}
\begin{table}[t]
  \caption{Number of parameters in each model.}
  \label{tab:parameters}
  \centering
  \begin{tabular}{c c c c}
    \toprule
    \textbf{System} & \textbf{StoRM (S)} & \textbf{GP-Unified (L)} & \textbf{Thunder (S/L)}\\
    \midrule
    Parameters & 55.6M & 106M & 27.8M/65.6M \\
    \bottomrule
  \end{tabular}
\end{table}

\section{Experiments}

\subsection{Experimental settings}

We benchmarked the performance of our proposed method (Thunder) on the VoiceBank + DEMAND dataset \cite{Voicebank, Demand}, consisting of 30 speakers from the Voicebank Corpus \cite{Voicebank}. We followed the prior works \cite{StoRM, SGMSE+} and separated the dataset into training (26 speakers), validation (speaker ``p226'', ``p287'') and testing (2 speakers) sets. The training and validation sets consist of 11,572 utterances corrupted by eight recorded noise samples from DEMAND and two artificially generated noise samples (babble and speech-shaped) at SNR levels of 0, 5, 10, and 15 dB, while the testing set contains 824 utterances, each contaminated with different noise samples at SNR levels of 2.5, 7.5, 12.5, and 17.5 dB. All speech data were sampled at 16 kHz.

We also followed prior works \cite{StoRM, SGMSE+, DiffusionJointPredictive} and used the Noise Conditional Score Network (NCSN++)\footnote{Our implementation was based on https://github.com/sp-uhh/storm} \cite{analysing_discriminative_diffusion} as a base architecture with 30 reverse diffusion steps for benchmarking with minor modifications. The model was used to predict clean speech instead of the score function, and the SDE was transformed into a Brownian bridge process. Note that the NCSN++ in StoRM \cite{StoRM} differs from that of SGMSE+ \cite{SGMSE+} and GP-Unified \cite{DiffusionJointPredictive} since it utilizes a smaller NCSN++ variant (27.8M) for both regression and diffusion models, whereas SGMSE+ and GP-Unified employ the larger variant (65.6M). For a fair comparison, we performed evaluations on both variants.

The model was trained for 100 epochs on one Nvidia RTX4090, with Adam optimizer, a learning rate of $2 \times 10^{-5}$, and a batch size of 8. We used Perceptual Evaluation of Speech Quality (PESQ) \cite{pesq}, Extended Short-Time Objective Intelligibility (ESTOI) \cite{estoi}, Scale-Invariant Signal-to-Distortion Ratio (SI-SDR) \cite{si_sdr}, and Scale-Invariant Signal-to-Artifact Ratio (SI-SAR) \cite{si_sdr} as evaluation metrics.


\setlength\tabcolsep{3pt}
\begin{table}[t!]
  \caption{The performance of Thunder (L) when varying the number of reverse time steps ($N$) using PC sampler. The RTF is the average time to process one second of audio. The experiments were conducted using Nvidia RTX 4090. Corrector denotes the corrector in the PC sampler.}
  \label{tab:rtf}
  \centering
  \begin{tabular}{c c c | c c}
    \toprule
     \textbf{N} & \textbf{Corrector} & \textbf{RTF[s]} $\downarrow$ & \textbf{PESQ} $\uparrow$ & \textbf{SI-SDR} $\uparrow$ \\
    \midrule
     30 & \xmark & 0.538 & 3.02 & 19.4\\
     30 & \cmark & 1.084 & 3.02 & 19.4\\
     15 & \xmark & 0.284 & 3.02 & 19.4\\
     15 & \cmark & 0.552 & 3.02 & 19.4\\
     1 & \xmark & 0.038 & 2.99 & 19.6\\
     1 & \cmark & 0.056 & 2.99 & 19.6\\
    \bottomrule
  \end{tabular}
\end{table}

\begin{figure}[t!]
  \centering
  \includegraphics[width=.95\linewidth]{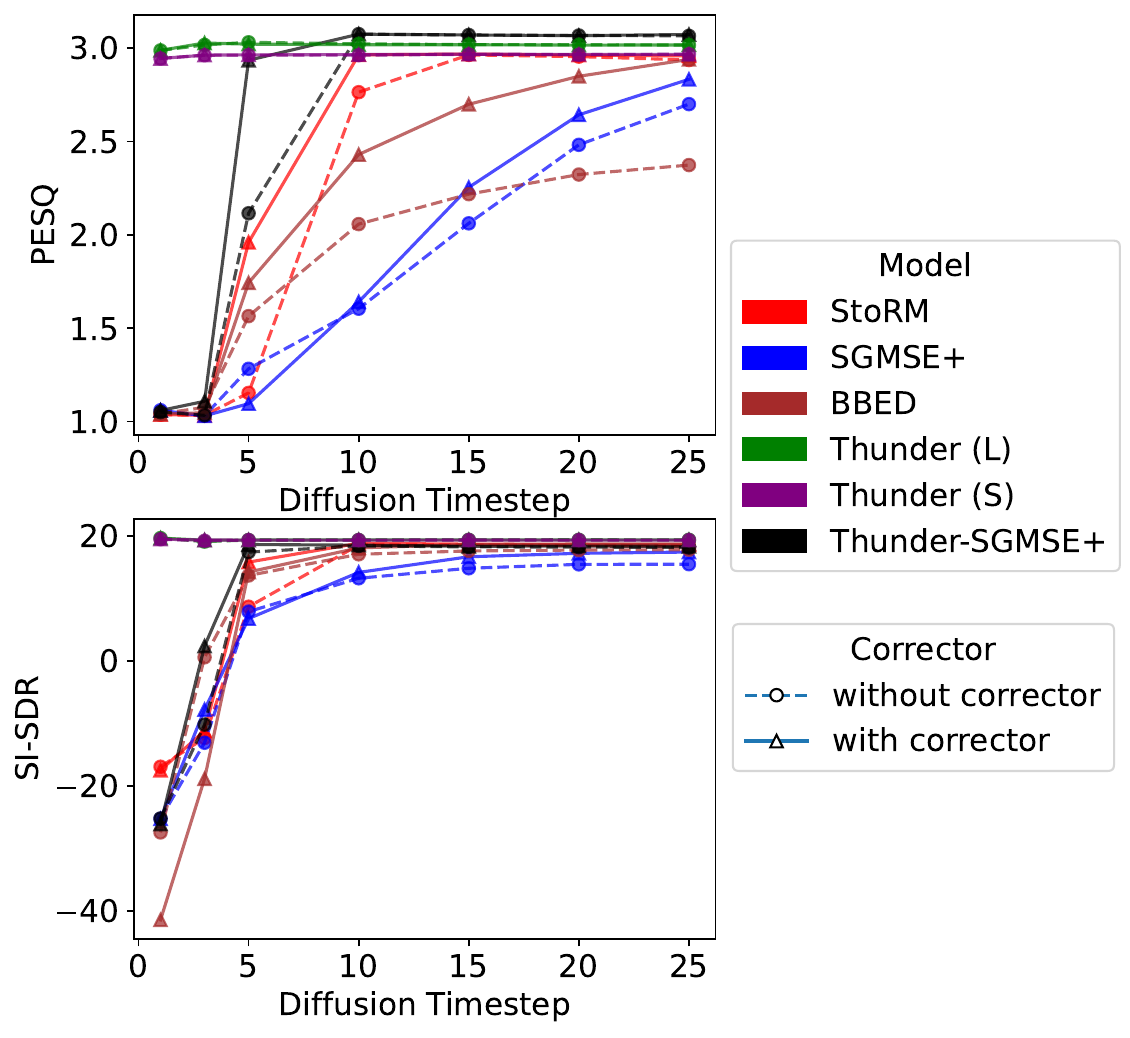}
  \caption{PESQ and SI-SDR under different numbers of diffusion time steps. Thunder performed competitively with other approaches even with just one diffusion step.}
  \label{fig:timestep}
\end{figure}

\subsection{In-domain evaluation}
Table \ref{tab:performance} compares our method to other approaches. It was found that our model achieved a competitive result compared to other state-of-the-art models.
  Additionally, compared to other diffusion-based approaches (``StoRM'', ``GP-Unified''), our method used half of the parameters (Table \ref{tab:parameters}) while performing competitively compared to StoRM and GP-Unified. 

  \vspace{-1cm}
  Figure \ref{fig:timestep} shows that our method maintains competitive results while requiring much fewer diffusion steps compared to the baselines (30 reverse steps), even achieving real-time inference (Table \ref{tab:rtf}). To justify our design choice, we also provide comparisons against BBDE \cite{ReducePrior} and Thunder-SGMSE+. The BBDE used the same SDE as ours, but it predicted the score function instead of the clean speech. On the other hand, Thunder-SGMSE+ is our method, but the SDE is changed to be the same as SGMSE+. Note that for $N=1$, the process requires two forward passes: regression and diffusion.

\subsection{Out-of-domain evaluation}
We further examine the generalizability of our method by performing an evaluation on the LibriFSD50k, the LibriSpeech dataset \cite{librispeech} corrupted by noise uniformly added from the FSD50k dataset \cite{fsd50k} at SNR levels ranging from 0 to 20, without any fine-tuning. The result in Table \ref{tab:generalized}  suggests that our model could still generalize under the out-of-domain setting, outperforming the other baselines (paired two-sample t-test, $p < 0.01$). Interestingly, there is only a slight degradation when reducing the number of reverse steps from thirty to one, implying that our regression mode is highly effective at eliminating the noise, requiring only one reverse step to refine.
\subsection{Effect of regression mode}
\begin{figure}[t!]
  \centering
  \includegraphics[width=.9\linewidth]{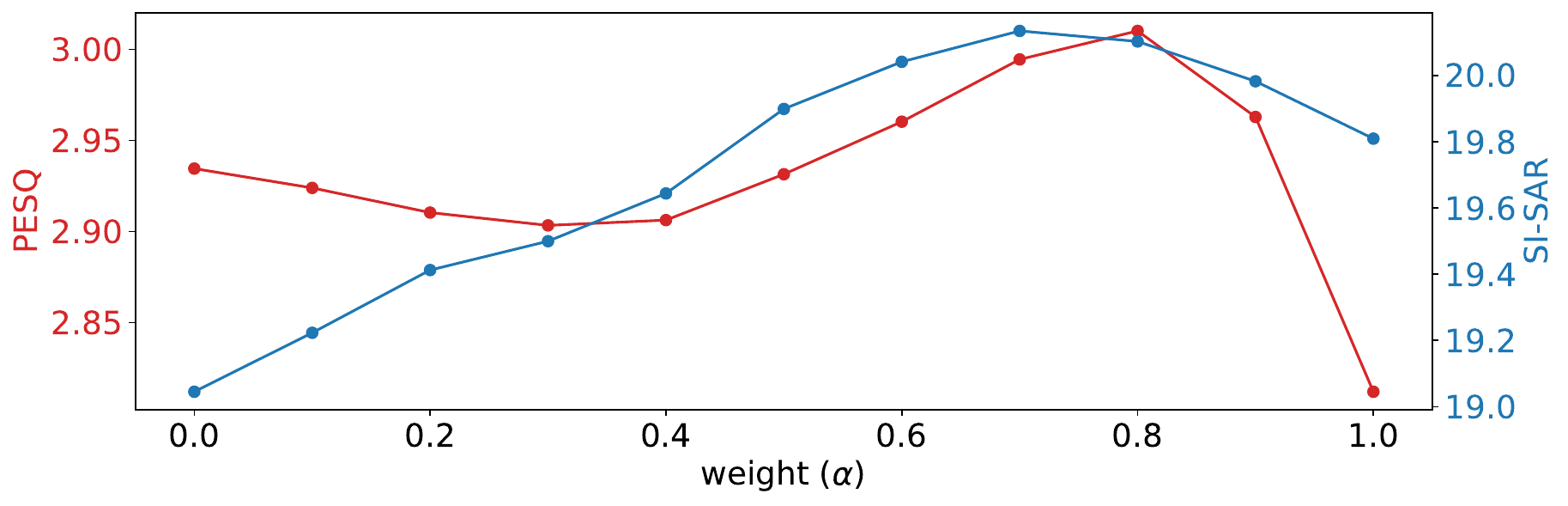}
   \caption{PESQ and SI-SAR of Thunder (L) at different interpolation weights $\alpha$. At high $\alpha$, performance degradation was observed due to artifacts from the regression mode, obstructing the refinement process during the diffusion mode.}
  \label{fig:blending_weight}
\end{figure}

We then investigated the effect of having a regression mode as the first step by varying the interpolation weight $\alpha$ from 0 to 1 (no blending with the original signal) while setting $t$ to 1. Figure \ref{fig:blending_weight} shows that high values of $\alpha$ led to audio quality degradation, as a sharp decline in PESQ and SI-SAR scores was observed when $\alpha > 0.8$ and $\alpha > 0.7$, respectively. This indicates that the regression mode generated excessive artifacts for the diffusion mode to refine. Despite this, the diffusion mode could still effectively eliminate artifacts when a sufficient degree of noisy speech $y$ was added to reduce the artifacts, thereby enhancing PESQ and improving the performance. On the other hand, the model yielded the lowest SI-SAR when the assistance from the regression mode ($\alpha=0$) was removed, suggesting its ability to reduce the difficulty of the reverse process, as also observed in an increase in SI-SAR and PESQ when $\alpha$ was around 0.5-0.8.

\setlength\tabcolsep{0.5pt}
\begin{table}[t!]
  \caption{The performance of Thunder (L) under mismatched training conditions on the FSD50k dataset. We achieved better generalization than the MetricGAN+ because a lower relative performance change between the out-of-domain and in-domain conditions was observed.}
  \label{tab:generalized}
  \centering
  \begin{tabular}{l c c c c}
    \toprule
     \textbf{System} & \textbf{Type} & \textbf{PESQ} $\uparrow$ & \textbf{SI-SDR} $\uparrow$ & \textbf{SI-SAR} $\uparrow$ \\
    \midrule
     Noisy & - & 1.92 & 10.0 & - \\
     \midrule
     MetricGAN+ & R & 2.18 & 5.8 & 6.1 \\
     NCSN++M (L) & R & 2.03 & 14.7 & 17.0 \\
     \midrule 
     SGMSE+ (L) (30 steps) & G & 2.19 & 14.2 & 15.8 \\
     StoRM (S) (30 steps) & R+G & 2.12 & 14.3 & 16.3 \\
    \midrule
    \textbf{Thunder (L)} & & & & \\
     Regression Mode & R & 2.04 & 14.7 & 16.7 \\
     Mixture (30 steps, $\alpha=0.8$) & R+G & 2.21 & 14.7 & 17.0 \\
     Mixture (1 step, $\alpha=0.8$) & R+G &  2.21 & 14.7 & 17.0 \\

    \bottomrule
  \end{tabular}
\end{table}

\section{Conclusion}
We proposed Thunder, a unified regression-diffusion model for speech enhancement. The model is trained to predict the clean speech instead of the score function to efficiently leverage the Brownian bridge process, allowing the model to possess both regressive and generative capabilities without incurring additional parameters. Our method achieves competitive results compared to other diffusion baselines on in-domain settings even with a single reverse diffusion step. It also outperforms other baselines in out-of-domain situations. For future work, we plan to extend Thunder to cover more general settings such as dereverberation.
\bibliographystyle{IEEEtran}
\bibliography{mybib}

\begin{thebibliography}{10}
\providecommand{\url}[1]{#1}
\csname url@samestyle\endcsname
\providecommand{\newblock}{\relax}
\providecommand{\bibinfo}[2]{#2}
\providecommand{\BIBentrySTDinterwordspacing}{\spaceskip=0pt\relax}
\providecommand{\BIBentryALTinterwordstretchfactor}{4}
\providecommand{\BIBentryALTinterwordspacing}{\spaceskip=\fontdimen2\font plus
\BIBentryALTinterwordstretchfactor\fontdimen3\font minus \fontdimen4\font\relax}
\providecommand{\BIBforeignlanguage}[2]{{%
\expandafter\ifx\csname l@#1\endcsname\relax
\typeout{** WARNING: IEEEtran.bst: No hyphenation pattern has been}%
\typeout{** loaded for the language `#1'. Using the pattern for}%
\typeout{** the default language instead.}%
\else
\language=\csname l@#1\endcsname
\fi
#2}}
\providecommand{\BIBdecl}{\relax}
\BIBdecl

\bibitem{app10176077}
G.~Park, W.~Cho, K.-S. Kim, and S.~Lee, ``{Speech Enhancement for Hearing Aids with Deep Learning on Environmental Noises},'' \emph{Applied Sciences}, vol.~10, no.~17, 2020.

\bibitem{9414346}
K.~Tan, X.~Zhang, and D.~Wang, ``{Real-Time Speech Enhancement for Mobile Communication Based on Dual-Channel Complex Spectral Mapping},'' in \emph{ICASSP}, 2021, pp. 6134--6138.

\bibitem{shon2019voiceid}
S.~Shon, H.~Tang, and J.~Glass, ``{VoiceID Loss: Speech Enhancement for Speaker Verification},'' in \emph{Proc. Interspeech}, 2019, pp. 2888--2892.

\bibitem{zhu2022joint}
Q.-S. Zhu, J.~Zhang, Z.-Q. Zhang, and L.-R. Dai, ``{A Joint Speech Enhancement and Self-Supervised Representation Learning Framework for Noise-Robust Speech Recognition},'' \emph{IEEE/ACM Transactions on Audio, Speech, and Language Processing}, vol.~31, pp. 1927--1939, 2023.

\bibitem{maas2012recurrent}
A.~Maas, Q.~V. Le, T.~M. O’Neil, O.~Vinyals, P.~Nguyen, and A.~Y. Ng, ``{Recurrent Neural Networks for Noise Reduction in Robust ASR},'' in \emph{Proc. Interspeech}, 2012.

\bibitem{koizumi2021snri}
Y.~Koizumi, S.~Karita, A.~Narayanan, S.~Panchapagesan, and M.~A.~U. Bacchiani, ``{SNRi Target Training for Joint Speech Enhancement and Recognition},'' in \emph{Proc. Interspeech}, 2022.

\bibitem{eskimez2018front}
S.~E. Eskimez, P.~Soufleris, Z.~Duan, and W.~Heinzelman, ``{Front-end speech enhancement for commercial speaker verification systems},'' \emph{Speech Communication}, vol.~99, pp. 101--113, 2018.

\bibitem{StoRM}
J.-M. Lemercier, J.~Richter, S.~Welker, and T.~Gerkmann, ``{StoRM: A Diffusion-Based Stochastic Regeneration Model for Speech Enhancement and Dereverberation},'' \emph{IEEE/ACM Transactions on Audio, Speech, and Language Processing}, vol.~31, pp. 2724--2737, 2023.

\bibitem{conv_tasnet}
Y.~Luo and N.~Mesgarani, ``{Conv-TasNet: Surpassing Ideal Time–Frequency Magnitude Masking for Speech Separation},'' \emph{IEEE/ACM Transactions on Audio, Speech, and Language Processing}, vol.~27, no.~8, pp. 1256--1266, 2019.

\bibitem{metricgan+}
S.-W. Fu, C.~Yu, T.-A. Hsieh, P.~Plantinga, M.~Ravanelli, X.~Lu, and Y.~Tsao, ``{MetricGAN+: An Improved Version of MetricGAN for Speech Enhancement},'' in \emph{Proc. Interspeech}, 2021, pp. 201--205.

\bibitem{gagnet}
A.~Li, C.~Zheng, L.~Zhang, and X.~Li, ``{Glance and gaze: A collaborative learning framework for single-channel speech enhancement},'' \emph{Applied Acoustics}, vol. 187, p. 108499, 2022.

\bibitem{segan}
S.~Pascual, A.~Bonafonte, and J.~Serrà, ``{SEGAN: Speech Enhancement Generative Adversarial Network},'' in \emph{Proc. Interspeech}, 2017, pp. 3642--3646.

\bibitem{CDiffSE}
Y.-J. Lu, Z.-Q. Wang, S.~Watanabe, A.~Richard, C.~Yu, and Y.~Tsao, ``{Conditional Diffusion Probabilistic Model for Speech Enhancement},'' in \emph{ICASSP}, 2022, pp. 7402--7406.

\bibitem{SGMSE+}
J.~Richter, S.~Welker, J.-M. Lemercier, B.~Lay, and T.~Gerkmann, ``{Speech Enhancement and Dereverberation With Diffusion-Based Generative Models},'' \emph{IEEE/ACM Transactions on Audio, Speech, and Language Processing}, vol.~31, pp. 2351--2364, 2023.

\bibitem{ScoreBasedDiffusion}
Y.~Song, J.~Sohl-Dickstein, D.~P. Kingma, A.~Kumar, S.~Ermon, and B.~Poole, ``{Score-Based Generative Modeling through Stochastic Differential Equations},'' in \emph{ICLR}, 2021.

\bibitem{DiffusionJointPredictive}
H.~Shi, K.~Shimada, M.~Hirano, T.~Shibuya, Y.~Koyama, Z.~Zhong, S.~Takahashi, T.~Kawahara, and Y.~Mitsufuji, ``Diffusion-based speech enhancement with joint generative and predictive decoders,'' in \emph{ICASSP}, 2024, pp. 12\,951--12\,955.

\bibitem{BrownianBridgeProcess}
I.~Karatzas and S.~E. Shreve, \emph{{Brownian Motion and Stochastic Calculus}}.\hskip 1em plus 0.5em minus 0.4em\relax Springer, 1998.

\bibitem{ScoreMatching}
Y.~Song and S.~Ermon, ``{Generative Modeling by Estimating Gradients of the Data Distribution},'' in \emph{Advances in Neural Information Processing Systems}, H.~Wallach, H.~Larochelle, A.~Beygelzimer, F.~d\textquotesingle Alch\'{e}-Buc, E.~Fox, and R.~Garnett, Eds., vol.~32.\hskip 1em plus 0.5em minus 0.4em\relax Curran Associates, Inc., 2019.

\bibitem{StochasticDifferentialBook}
S.~S{\"a}rkk{\"a} and A.~Solin, \emph{{Applied stochastic differential equations}}.\hskip 1em plus 0.5em minus 0.4em\relax Cambridge University Press, 2019, vol.~10.

\bibitem{ReducePrior}
B.~Lay, S.~Welker, J.~Richter, and T.~Gerkmann, ``{Reducing the Prior Mismatch of Stochastic Differential Equations for Diffusion-based Speech Enhancement},'' in \emph{Proc. Interspeech}, 2023, pp. 3809--3813.

\bibitem{vae}
D.~P. Kingma and M.~Welling, ``{Auto-Encoding Variational Bayes},'' in \emph{ICLR}, 2014.

\bibitem{ArtifactsASR}
K.~Iwamoto, T.~Ochiai, M.~Delcroix, R.~Ikeshita, H.~Sato, S.~Araki, and S.~Katagiri, ``{How bad are artifacts?: Analyzing the impact of speech enhancement errors on ASR},'' in \emph{Proc. Interspeech}, 2022, pp. 5418--5422.

\bibitem{SGMSE}
S.~Welker, J.~Richter, and T.~Gerkmann, ``{Speech Enhancement with Score-Based Generative Models in the Complex STFT Domain},'' in \emph{Proc. Interspeech}, 2022, pp. 2928--2932.

\bibitem{analysing_discriminative_diffusion}
J.-M. Lemercier, J.~Richter, S.~Welker, and T.~Gerkmann, ``Analysing diffusion-based generative approaches versus discriminative approaches for speech restoration,'' in \emph{ICASSP}, 2023, pp. 1--5.

\bibitem{Voicebank}
V.-B. Cassia, W.~Xin, T.~Shinji, and Y.~Junichi, ``{Investigating RNN-based speech enhancement methods for noise-robust Text-to-Speech},'' \emph{9th ISCA Workshop on Speech Synthesis Workshop (SSW 9)}, 09 2016.

\bibitem{Demand}
J.~Thiemann, N.~Ito, and E.~Vincent, ``{The Diverse Environments Multi-channel Acoustic Noise Database (DEMAND): A database of multichannel environmental noise recordings},'' \emph{Proceedings of Meetings on Acoustics}, vol.~19, no.~1, p. 035081, 05 2013.

\bibitem{pesq}
P.~Recommendation, ``{Application Guide for objective quality measurement based on recommendation},'' \emph{ITUt}, vol. 862, p. 862, 2005.

\bibitem{estoi}
J.~Jensen and C.~H. Taal, ``{An Algorithm for Predicting the Intelligibility of Speech Masked by Modulated Noise Maskers},'' \emph{IEEE/ACM Transactions on Audio, Speech, and Language Processing}, vol.~24, no.~11, pp. 2009--2022, 2016.

\bibitem{si_sdr}
J.~L. Roux, S.~Wisdom, H.~Erdogan, and J.~R. Hershey, ``{SDR – Half-baked or Well Done?}'' in \emph{ICASSP}, 2019, pp. 626--630.

\bibitem{librispeech}
V.~Panayotov, G.~Chen, D.~Povey, and S.~Khudanpur, ``Librispeech: An asr corpus based on public domain audio books,'' in \emph{ICASSP}, 2015, pp. 5206--5210.

\bibitem{fsd50k}
E.~Fonseca, X.~Favory, J.~Pons, F.~Font, and X.~Serra, ``Fsd50k: An open dataset of human-labeled sound events,'' \emph{IEEE/ACM Transactions on Audio, Speech, and Language Processing}, vol.~30, pp. 829--852, 2022.

\end{thebibliography}

\end{document}